\documentclass{article}
\usepackage{frascatiphys,here,graphicx,subfigure}
\newcommand{\BR}{\mathcal{B}}

\newcommand{\psp}{\psi(2S)}
\newcommand{\psip}{\psi(2S)}

\newcommand{\jpsi}{J/\psi}

\newcommand{\ppjpsi}{\pi^+\pi^-J/\psi}

\newcommand{\chicz}{\chi_{c0}}
\newcommand{\chico}{\chi_{c1}}
\newcommand{\chict}{\chi_{c2}}
\newcommand{\chicJ}{\chi_{cJ}}

\newcommand{\EE}{e^+e^-}
\newcommand{\MM}{\mu^+\mu^-}

\newcommand{\piz}{\pi^0}

\newcommand{\ppb}{p\overline{p}}
\newcommand{\aab}{\Lambda\overline{\Lambda}}

\newcommand{\ppbar}{p \overline{p}}

\newcommand{\g}{\gamma}

\newcommand{\gpppr}{\gamma \pi^+\pi^-p\bar{p}}
\newcommand{\gkkkk}{\gamma K^+K^-K^+K^-}

\newcommand{\pipi}{\pi^+\pi^-}
\newcommand{\kskp}{K^0_S K^+ \pi^- + c.c.}
\newcommand{\kk}{K^+K^-}
\newcommand{\beq}{\begin{equation}}
\newcommand{\eeq}{\end{equation}}
\newcommand{\beqn}{\begin{eqnarray}}
\newcommand{\eeqn}{\end{eqnarray}}
\newcommand{\beqns}{\begin{eqnarray*}}
\newcommand{\eeqns}{\end{eqnarray*}}
\newcommand{\bfg}{\begin{figure}}
\newcommand{\efg}{\end{figure}}
\newcommand{\bitm}{\begin{itemize}}
\newcommand{\eitm}{\end{itemize}}
\newcommand{\bnum}{\begin{enumerate}}
\newcommand{\enum}{\end{enumerate}}
\newcommand{\btbl}{\begin{table}}
\newcommand{\etbl}{\end{table}}
\newcommand{\btbu}{\begin{tabular}}
\newcommand{\etbu}{\end{tabular}}
\begin{document}
\title{Recent BES Results on Charmonium Decays}
\author{
Chang-Zheng Yuan      \\
(for the BES Collaboration)\\
{\em Institute of High Energy Physics, Beijing 100049, China} }
\maketitle \baselineskip=11.6pt

\begin{abstract}
In this talk, we present the recent results on charmonium decays
from the BES experiment at the BEPC collider. The analyses are
based on a 14~million $\psp$ events data sample. We report results
on leptonic decays, hadronic decays, and radiative decays of
$\psp$, as well as hadronic decays of $\chicJ$ states and rare or
forbidden decays of $\jpsi$.
\end{abstract}
\baselineskip=14pt

\section{Introduction}

We report the recent analyses on charmonium decays with the $\psp$
data collected with the BESII detector~\cite{bes} at the BEPC
collider. The data sample has 14~million produced $\psp$
events~\cite{moxh}.

\section{\boldmath Branching fraction of $\psp\to \tau^+\tau^-$}

The $\psi(2S)$ data provides an opportunity to compare the
coupling of the photon to the three generation leptons by studying
the leptonic decays $\psi(2S)\to e^+e^-$, $\mu^+\mu^-$, and
$\tau^+\tau^-$. The leptonic decay branching fractions are
described by the relation $B_{ee}\simeq B_{\mu\mu}\simeq
B_{\tau\tau}/0.3885$, which are in good agreement with BESI
measurement~\cite{besitt}. The branching fraction for
$\psi(2S)\to\tau^+\tau^-$ is remeasured~\cite{bestaotao} with
$\tau^+ \tau^-$ pair reconstructed with their pure leptonic
decays. At $\psi(2S)$ resonance, 1015 signal events are observed,
and the QED process contributes 516 events measured with a data
sample at $\sqrt s=3.65$~GeV. The branching fraction is calculated
to be $(0.310\pm 0.021\pm 0.038)\%$, where the first error is
statistical and the second systematic. This improves the precision
and the $e-\mu-\tau$ universality is tested at a higher level than
at BESI.

\section{\boldmath $\psp$ radiative decays}

Besides conventional meson and baryon states, QCD also predicts a
rich spectrum of glueballs, hybrids, and multi-quark states in the
1.0 to 2.5~$\hbox{GeV}/c^2$ mass region. Therefore, searches for
the evidence of these exotic states play an important role in
testing QCD. The radiative decays of $\psip$ to hadrons are
expected to contribute about 1\% to the total $\psip$ decay
width~\cite{PRD_wangp}. However, the measured channels only sum up
to about 0.05\%~\cite{PDG}.

\begin{figure}\centering
\includegraphics[height=12.0cm]{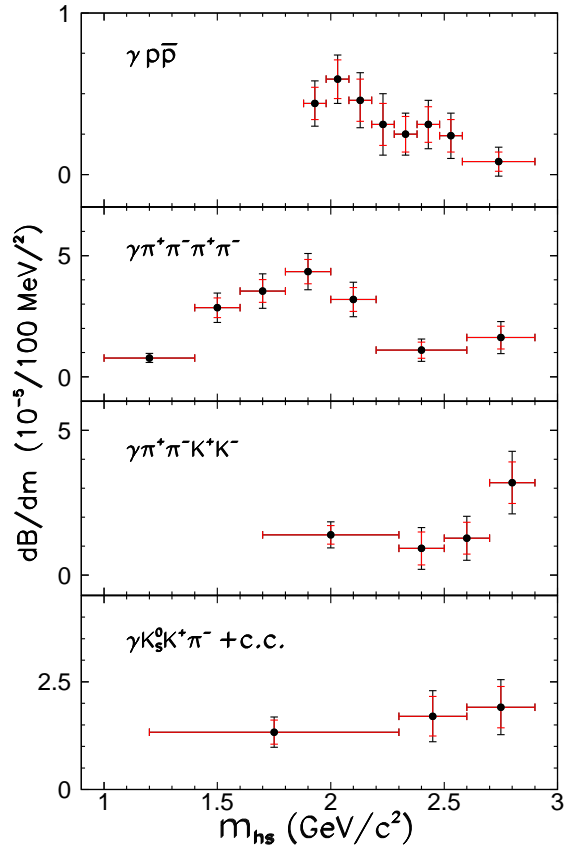}
\caption{ \label{difbr} Differential branching fractions for
$\psip$ decays into $\gamma\ppbar$, $\gamma 2(\pipi)$, $\gamma K^+
K^- \pipi$, and $\gamma \kskp$ Here $m_{hs}$ is the invariant mass
of the hadrons in each final state. For each point, the smaller
longitudinal error is the statistical error, while the bigger one
is the quadratic sum of statistical and systematic errors. }
\end{figure}

We measured the decays of $\psip$ into $\gamma\ppb$, $\gamma
2(\pipi)$, $\gamma \kskp$, $\gamma K^+ K^- \pipi$, $\gamma
K^{*0}K^-\pi^+ +c.c.$, $\gamma K^{*0}\bar K^{*0}$,
$\gamma\pipi\ppb$, $\g2(\kk)$, $\gamma3(\pipi)$, and $\gamma
2(\pi^+\pi^-)K^+K^-$, with the invariant mass of the hadrons
($m_{hs}$) less than 2.9~$\hbox{GeV}/c^2$ for each decay
mode~\cite{bes2rad}. The differential branching fractions are
shown in Fig.~\ref{difbr}. The branching fractions below
$m_{hs}<2.9$~$\textrm{GeV}/c^2$ are given in Table~\ref{Tot-nev},
which sum up to $0.26\%$ of the total $\psip$ decay width. We also
analyzed $\psp\to \gamma\pipi$ and $\gamma\kk$ modes to study the
resonances in $\pipi$ and $\kk$ invariant mass spectrum.
Significant signals for $f_2(1270)$ and $f_0(1710)$ were observed,
but the low statistics prevent us from drawing solid conclusion on
the other resonances~\cite{agnes}.

\begin{table}
\caption{\label{Tot-nev} Branching fractions for $\psip\to\gamma
+hadrons$ with $m_{hs}<2.9$ $\hbox{GeV}/c^2$, where the upper
limits are determined at the 90\% C.L.}
\begin{center}
\begin{tabular}{ll} \hline \hline
Mode & $\BR(\times 10^{-5})$\\\hline
$\gamma p\bar{p}$ & 2.9$\pm$0.4$\pm$0.4 \\
$\gamma 2(\pi^+\pi^-)$ & 39.6$\pm$2.8$\pm$5.0\\
$\gamma K^0_S K^+\pi^-+c.c.$  & 25.6$\pm$3.6$\pm$3.6 \\
$\gamma K^+ K^-\pi^+\pi^-$ & 19.1$\pm$2.7$\pm$4.3 \\
$\gamma K^{*0} K^+\pi^-+c.c.$& 37.0$\pm$6.1$\pm$7.2\\
$\gamma K^{*0}\bar K^{*0}$&$24.0\pm 4.5\pm 5.0$\\
$\gpppr$& 2.8$\pm$1.2$\pm$0.7 \\
$\gkkkk$ &  $<4$\\
$\gamma3(\pipi)$&  $<17$\\
$\gamma2(\pi^+\pi^-)K^+K^-$& $<22$ \\
\hline \hline
\end {tabular}
\end{center}
\end{table}

\section{\boldmath $\psp$ hadronic decays}

\subsection{$\sigma$ in $\psp\to \ppjpsi$}

The process $\psi(2S)\rightarrow \pi^+\pi^-J/\psi$,
$J/\psi\rightarrow \mu^+\mu^-$ is analyzed to study the $\pipi$
interaction~\cite{lig}.

We fit the data with two different models. For the first model,
using four different Breit-Wigner parameterizations, the data can
be well fitted with a $\sigma$ term and a contact term. The final
best estimate of the $\sigma$ pole position is
$(552^{+~84}_{-106})-i(232^{+81}_{-72})$~MeV$/c^2$, where the
errors cover the statistical and systematic errors, including the
differences in the Breit-Wigner parameterizations.

We also fit our data according to the scheme in Ref.~\cite{guofk}.
It is found that {the $\pi\pi$ S-wave FSI plays a dominant role in
$\psi(2S)\rightarrow\pi^+\pi^-J/\psi$, while the contribution from
the contact term is small. The $\sigma$ pole used in this fit,
$469-i203$~MeV/$c^2$ is consistent with the fits to the
Breit-Wigner functions. This implies that, although the two
theoretical schemes are very different, both of them find the
$\sigma$ meson at similar pole positions.

If the $\sigma$ meson exists, the pole should occur universally in
all $\pi\pi$ system with correct quantum numbers. Our analysis
demonstrates that, in $\psi(2S)\rightarrow\pi^+\pi^-J/\psi$, one
can still determine the pole location in good agreement with that
obtained from $J/\psi\rightarrow \omega\pi^+\pi^-$ decay
{~\cite{ref_41}}.

\subsection{Hadronic decays with Baryons in the final states}

In perturbative QCD (pQCD), hadronic decays of both $\psi(2S)$ and
$J/\psi$ proceed dominantly via an annihilation of $c\bar c$
quarks into three gluons or one photon, followed by a
hadronization process. This yields the so-called ``12\% rule'',
{\it i.e.} $Q_h\equiv {B_{\psip\to h}\over B_{J/\psi\to
h}}={B_{\psip\to \EE}\over B_{J/\psi\to \EE}}\simeq 12\%$.
Table~\ref{hadbr} summarizes recent measurements on $\psip$ decays
at BES. For a number of $\psip$ decays $Q_h$s are in agreement
with $12\%$ within $1\sim 2\sigma$.

The branching fractions of $\psip$ decays into octet baryon are
measured~\cite{jiaojb} and listed in Table~\ref{hadbr}. For
$\psip\to N\bar N\pi$~\cite{pppi0,pnpi}, the ratio of the measured
branching fractions is
$\mathcal{B}(\psip\to\ppbar\pi^0):\mathcal{B}(\psip\to p\bar
n\pi^-):\mathcal{B}(\psip\to\bar pp\pi^+) = 1:1.86\pm0.27:
1.91\pm0.27$, which is consistent with the isospin symmetry
prediction $1:2:2$.

\begin{table}
\caption{\label{hadbr} Branching fractions for $\psip$ hadronic
decays. Here $Q_h$ is defined as $Q_h={\BR(\psip\to h)\over
\BR(J/\psi\to h)}$, where $\BR(J/\psi\to h)$s are taken from
\cite{PDG}.}
\begin{center}
\begin{tabular}{cccccc} \hline \hline
Mode: $h$  &  $\BR(\times 10^{-4})$ &$Q_h$(\%)\\
\hline
$\ppbar$&$3.36\pm0.09\pm0.25$&$14.9\pm1.4$\\
$\Lambda\bar\Lambda$&$3.39\pm0.20\pm0.32$&$16.7\pm2.1$\\
$\Sigma^0\bar\Sigma^0$&$2.35\pm0.36\pm0.32$&$16.8\pm3.6$\\
$\Xi^-\Xi^+$&$3.03\pm0.40\pm0.32$&$16.8\pm4.7$\\
$\ppbar\piz$ & $1.32\pm 0.10\pm 0.15$ & $12.1\pm 1.9$ \\
$p\bar n\pi^-$&$2.45\pm0.11\pm0.21$&$12.0\pm1.5$\\
$\bar pn\pi^+$&$2.52\pm0.12\pm0.22$&$12.9\pm1.7$ \\
 \hline \hline
\end {tabular}
\end{center}
\end{table}

No $\psp\to \aab\piz$ and $\aab\eta$ are observed and the upper
limits on the production rates are determined~\cite{bes2llp}. We
also measure these two modes in $\jpsi$ decays. In our analysis,
it is found that $\aab\piz$ is seriously contaminated by
$\jpsi\to\Sigma^0\pi^0 \bar\Lambda+c.c.$ and
$\Sigma^+\pi^-\bar\Lambda+c.c.$ After removing these backgrounds,
no significant signal is observed for
$\jpsi\to\Lambda\bar\Lambda\pi^0$, and the upper limit is
determined to be $\BR(\jpsi\to\Lambda\bar\Lambda\pi^0)<0.64\times
10^{-4}$ at the 90\% C.L.; while the branching fraction of
$\jpsi\to \Lambda\bar\Lambda\eta$ is determined to be
$(2.62\pm0.60\pm0.44)\times 10^{-4}$. This indicates that
$\jpsi\to\Lambda\bar\Lambda\pi^0$ is suppressed due to the isospin
conservation, and the previous measurements by DM2~\cite{dm2} and
BESI~\cite{bes1} underestimate the background contribution.

\section{\boldmath $\chicJ\to {\rm three~pseudoscalars}$}

Decays of $\chicz$ and $\chict$ into three pseudoscalars are
suppressed by the spin-parity selection rule. We measured the
branching fractions of $\chico$ decays into $\kskp$ and
$\eta\pipi$ and intermediate states involved~\cite{mppbar4}.

$\kskp$ events are mainly produced via $K^*(892)$ intermediate
state, and $\eta\pipi$ events via $f_2(1270)\eta$ and
$a_0(980)\pi$. The branching fractions with these resonances are
\begin{eqnarray*}
\BR(\chico\to K^*(892)^0\bar
K^0+c.c.)=(1.1\pm0.4\pm0.1)\times 10^{-3},\nonumber\\
\BR(\chico\to K^*(892)^+
K^-+c.c.)=(1.6\pm0.7\pm0.2)\times 10^{-3},\nonumber\\
\BR(\chico\to f_2(1270)\eta)=(3.0\pm0.7\pm0.5)\times
10^{-3},\nonumber\\
\BR(\chico\to a_0(980)^+\pi^-+c.c.\to
\eta\pipi)=(2.0\pm0.5\pm0.5)\times 10^{-3}.
\end{eqnarray*}
Except for $\chico\to \kskp$, all other modes are the first
observations.

\section{\boldmath Search for rare and forbidden decays}

\subsection{Upper limit on $\BR(\jpsi\to\gamma\gamma)$}

We searched for the C-parity violating decay, $\jpsi\to
\gamma\gamma$~\cite{wangzy_gg}. In a previous
measurement~\cite{cntr}, $\jpsi$ produced directly in $\EE$
annihilation was used, and the upper limit measured is
$\BR(\jpsi\to\gamma\gamma)<5\times 10^{-4}$ at 90\% C.L. In our
analysis we studied this decay via $\psip\to\ppjpsi$,
$\jpsi\to\gamma\gamma$. Therefore, the QED background is strongly
suppressed since we observe a $\pipi$ pair plus two photons and do
not base our search just on $\gamma\gamma$ invariant mass
distribution.

The total number of events in the signal region is 52, the peaking
background is 30.4 and the smooth background is 18.6. With the
Bayesian method, the upper limit on the number of $\jpsi\to
\gamma\gamma$ events is estimated to be 16 at the 90\% C.L., in
which the systematic errors have been taken into account.
Therefore, the upper limit on ${\cal B}(\jpsi\to\gamma\gamma)$ is
measured to be $2.2\times 10^{-5}$. Our upper limit for the
C-violating decay is about 20 times more stringent than the
previous measurement. It indicates that there is no obvious
C-parity violation.

\subsection{Search for $\jpsi$ decays into invisible particles}

Invisible decays of quarkonium states such as $\jpsi$ and
$\Upsilon$ offer a window into what may lie beyond the standard
model (SM)~\cite{0}.

In order to detect invisible $\jpsi$ decay, we use
$\psip\to\ppjpsi$ and infer the presence of the $\jpsi$ resonance
from the $\jpsi$ peak in the distribution of mass recoiling
against the $\pi^+\pi^-$~\cite{wangzy_inv}. A $\chi^2$ fit is used
to extract the number of $\jpsi$ events in the $\pipi$ recoiling
mass distribution in the range
$3.0$~GeV/$c^2<M^{\mbox{recoil}}_{\pipi}< 3.2$~GeV/$c^2$. The
function to describe the signal comes from the shape of the
$\pipi$ recoiling mass spectrum from the control sample $\psip\to
\ppjpsi$, $\jpsi\to \MM$. The fit yields $6424\pm 137$ events,
which includes the contributions from both signal and peaking
backgrounds, since they have the same probability density
functions in the fit. After subtracting the expected backgrounds
from the fitted yields, we get the number of events of $\psip\to
\ppjpsi$, $\jpsi\to \mbox{invisible} $ to be $406\pm 385$. The
upper limit is determined to be $N^{\jpsi}_{UL}=1045$ at the 90\%
C.L. from the Feldman-cousins frequentist approach. The upper
limit on the ratio $\frac{{\cal B}(\jpsi\to
\mbox{invisible})}{{\cal B}(\jpsi\to \MM)}$ is $1.0\times 10^{-2}$
at 90\% C.L. This measurement improves by a factor of 3.5 the
bound on the product of the coupling of the $U$ boson to the
$c$-quark and LDM particles as described in Ref.~\cite{0}.

\section{\boldmath Summary}

Using the 14~M $\psip$ events sample taken with the BESII detector
at the BEPC storage ring, BES experiment provided many interesting
results in charmonium decays, including the observation of many
$\psip$ radiative decays, some $\psip$ hadronic decays, $\chicJ$
decays, and the rare and forbidden $\jpsi$ decays. These results
shed light on the understanding of SM.

\section{Acknowledgements}

We thanks BES colleagues for the nice work presented in this talk.
This work was supported in part by National Natural Science
Foundation of China under Contract No.~10491303.

\end{document}